\documentclass[english,11pt]{article}
          \usepackage{babel}
          \usepackage[T1]{fontenc}
          \usepackage[latin2]{inputenc}
          \usepackage{pslatex}
          \usepackage{epsf}
\def\refitem#1\par{\parindent=0truemm\parskip=0pt\hangindent=2truecm
     \frenchspacing#1\par}

\def \AA #1 #2 {{\em Astron.Astrophys.\/} {\bf #1}, #2.}
\def \ApJ #1 #2 {{\em Astrophys.J.\/} {\bf #1}, #2.}
\def \Acta #1 #2 {{\em Acta Astr.\/} {\bf #1}, #2.}
\def \AJ #1 #2 {{\em Astron.J.\/} {\bf #1}, #2.}
\def \PASP #1 #2 {{\em Publ.Astr.Soc.Pacific\/} {\bf #1}, #2.}
\def \MNRAS #1 #2 {{\em M.N.R.A.S.\/} {\bf #1}, #2.}
\def \PASJ #1 #2 {{\em Publ.Astr.Soc.Japan\/} {\bf #1}, #2.}
\def \BA #1 #2 {{\em Baltic Astronomy\/} {\bf #1}, #2.}

\begin{document}

\vskip 30 truemm

\title{\bf On the Amplitudes of Superhumps }

\author{ J. Smak } 
\date{ Nicolaus Copernicus Astronomical Center,~Polish Academy of 
Sciences \\ ul.~Bartycka~18, 00-716~Warszawa, Poland \\
{ e-mail: jis@camk.edu.pl} }

\maketitle

\vskip 15 truemm

\begin{abstract}

Amplitudes of bolometric light curves produced by 2D and 3D SPH simulations 
are used to determine the corresponding visual amplitudes. They turn 
out to be $\sim 10$ times lower than typical amplitudes of superhumps. 
This means a major failure of the tidal model of superhumps. 

\vskip 10 truemm


\end{abstract}

\section { Introduction }

The name "superhumps" refers to periodic light variations with periods 
slightly longer than the orbital periods which are observed in dwarf 
novae during their superoutbursts and in the so-called permanent 
superhumpers (cf. Warner 1995, Hellier 2001 and references therein). 
Their amplitudes are -- typically -- 0.3 mag. 

Superhumps are commonly believed to be due to the tidal effects in the 
outer parts of accretion disks leading -- via the 3:1 resonance -- to 
the formation of an eccentric outer ring undergoing apsidal motion. 
Following the pioneering work by Whitehurst (1988) and Hirose and Osaki 
(1990) numerous authors published results of their 2D and 3D SPH 
simulations, their main goal being to reproduce the superhump periods. 

Less attention was paid to the problem of amplitudes. In fact, there 
have been only few papers (for references -- see Section 3) presenting 
light curves resulting from such simulations. 
Their amplitudes, compared with the observed superhump amplitudes, 
could provide another crucial test for the tidal model. 
Unfortunately, such a direct comparison has been -- so far -- impossible 
for two reasons. 
First, because the model light curves are bolometric. Secondly, because 
in nearly all cases they do not refer to the full disk 
but are calculated with respect to the bolometric luminosity of the 
outer part of the disk (their amplitudes being, obviously, larger 
than the true amplitudes). 
The only exception were the visual, full disk light curves published  
by Simpson et al. (1998, Fig.2). Their amplitudes were very small 
($\sim 0.04$ mag.) but this could be attributed to the fact that model 
parameters used in those simulation were chosen to represent AM CVn.  

To clarify this problem we use a simple method (Section 2) to determine 
(Section 3) the full disk visual amplitudes corresponding to the amplitudes 
of "partial" bolometric light curves produced by 2D and 3D SPH simulations. Results will be summarized in Section 4.

% ***section 2
\section { Definitions and Formulae }

To begin with we define the amplitude as 

%***Eq.1
\begin{equation}
{A~=~{ {L_{max}-L_{min}}\over L_{min} }~
   =~{ {L_{max}}\over {L_{min}} }~-~1 }~.       
\end{equation}

\noindent
The amplitudes referring to the full disk will be designated as "$A$", 
while those referring only to its outer parts (see below) -- as "$a$". 

In what follows we will use the amplitudes of bolometric light curves 
of superhumps resulting from 2D or 3D numerical SPH simulations made 
with various values of the mass ratio $q$. 
Important for further analysis is the fact that in nearly all cases those 
light curves do not refer to the full disk but were calculated using 
only the outer parts of the disk with $R_i<R<R_d$ (the resulting amplitudes $a_{bol}$ being, obviously, larger than the true amplitudes $A_{bol}$). 
Accordingly, the luminosities referring to the full disk will be 
designated as $L_{bol}$ and $L_{V}$, while those of the outer 
part -- as $L_{bol,i}$ and $L_{V,i}$. 

The superhump light source (SLS) is located in the outer parts of 
the disk and covers only small fraction $x$ of its total area. 
From "dissipation maps" (e.g. Fig.8 of Murray 1998) we estimate that 
at superhump maximum $x \approx 0.05$. 
For the specific location of SLS we assume that it is located in the 
ring between $R_{sh}$ and $R_{sh}+\Delta R$, where $\Delta R=0.1R_d$. 
The fraction of the ring area covered by SLS is then 
$x_r=x/(2r\Delta r+\Delta r^2)$, where $\Delta r=\Delta R/R_d$. 
The corresponding ring luminosities will be designated as 
$L_{bol,r}$ and $L_{V,r}$.   
To see how our results depend on the adopted value of $R_{sh}$ we used 
$R_{sh}/R_d=0.6$, 0.7, 0.8, and 0.9. In all cases the resulting visual 
amplitudes (see below) turned out to be very similar. 

System parameters used in our calculations are: 
the mass of the primary, for which we adopt $M_1=0.8M\odot$ -- typical 
for dwarf dwarf novae of the SU UMa type, and the mass ratio $q$, 
which is already specified for each model. 
Together with the Kepler Law and the mass-radius relation for the 
secondary they uniquely determine all other parameters.  
Needed in further calculations are: the radius of the disk, calculated as $R_d=R_{tid}=0.9R_{Roche}$, and the full disk area: $S_d=2\pi R_d^2$ (where 
"2" refers to the two sides of the disk). 
The bolometric and visual luminosities at minimum -- 
$L_{bol},~L_{bol,i},~L_{bol,r}$ and $L_V,~L_{V,i},~L_{V,r}$ -- are then calculated using the standard formula for $T_e(R)$ applicable to the case 
of steady-state accretion:  

%***Eq.2
\begin{equation}
{ \sigma T_e^4~=~{3\over{8\pi}}~{{GM_1}\over{R^3}}~\dot M~
     [1~-~(R_1/R)^{1/2}] }~.
\end{equation}

\noindent
For the accretion rate we adopt $\log \dot M=17.5$, obtained 
from recent analysis of superoutbursts of Z Cha (Smak 2008). 
Worth adding is that our particular choice of $M_1$ and $\dot M$ has only 
little effect on results discussed below. 

We now assume that the SLS can be represented as a region of higher 
temperature $T_{sh}=$const., covering area $x S_d$.   
If so, the bolometric luminosity of the outer part of the disk at maximum (including SLS) can be written as

%***Eq.3
\begin{equation}
{ L_{bol,i}^{max}~=~(1~+~a_{bol})~ L_{bol,i}~=~
       L_{bol,i}~-x_r~ L_{bol,r}~+~x~ S_d~ \sigma T_{sh}^4 }~.
\end{equation}

\noindent
This equation can be used to determine $T_{sh}$. 
Turning to the visual luminosity of the full disk at maximum (also 
including SLS) we can write 

%***Eq.4
\begin{equation}
{ L_V^{max}~=~L_V~-~x_r~ L_{V,r}~+~x~ S_d~ f_V(T_{sh})}~, 
\end{equation}

\noindent
where $f_V(T)$ is the visual flux (per unit area). 
The resulting visual amplitude is 

%***Eq.5
\begin{equation}
{A_V~=~{ {L_V^{max}}\over {L_V} }~-~1 }~.       
\end{equation}

% ***section 3
\section { The Visual Amplitudes }

We begin with models based on 2D calculations. As demonstrated by 
Smith et al. (2007) such models are insufficient to produce reliable 
superhump periods and their amplitudes. 
In particular, the comparison of light curves presented in their Fig.6 
shows that the amplitudes resulting from 2D simulations are 2-4 times 
larger than those obtained from 3D calculations. 
Furthermore, the 2D light curves often have peculiar shapes including 
narrow peaks giving amplitudes much larger than those corresponding 
to the smooth part of the light curve.

% ***Table 1
\begin{table}[h!]
\parskip=0truept
\baselineskip=0pt
\bigskip
\centerline{Table 1}
\medskip
\centerline{ Amplitudes of Superhumps from 2D Simulations }
\medskip
$$\offinterlineskip \tabskip=0pt
\vbox {\halign {\strut
\vrule width 0.3truemm #&     %1
\quad#\quad&		      %2  mod
\vrule#&		      %3
\quad\hfil#\hfil\quad&	      %6  q
\vrule#&		      %7
\quad\hfil#\hfil\quad&	      %8  Ri/A
\vrule#&		      %9
\hfil#\hfil&	    %10 Abol
\vrule#&		      %11
\hfil#\hfil&	    %12 Av
\vrule width 0.3 truemm # \cr %13
\noalign {\hrule height 0.3truemm}
&&&&&&&&&&\cr
& Model/Run \hfil && $q$ && $R_i/A$ && $a_{bol}$ && $A_V$ &\cr
&&&&&&&&&&\cr
\noalign {\hrule}
&&&&&&&&&&\cr
&Hirose and Osaki (1990)\hfil && 0.150 && 0.25 &&~~2.2:/5.8: &&~~0.12:/0.20:&\cr
&Murray (1996)		\hfil && 0.176 && $^*$)&& 0.08 && 0.080 &\cr
&			      && 0.176 && 0.05 && 0.14 && 0.077 &\cr
&Murray (1998) \hfil  - 2 && 0.250 && 0.05 && 0.04 && 0.043 &\cr
&\hfil		      - 3 && 0.250 && 0.05 &&~~0.09/0.30 &&~~0.067/0.121 &\cr
&\hfil		      - 4 && 0.176 && 0.05 &&~~0.09/0.32 &&~~0.061/0.120 &\cr
&\hfil		      - 5 && 0.176 && 0.05 && 0.09 && 0.061 &\cr
&\hfil		      - 6 && 0.176 && 0.05 && 0.12 && 0.071 &\cr
&\hfil		      - 8 && 0.176 && 0.05 && 0.21 && 0.096 &\cr
&\hfil		      - 9 && 0.111 && 0.05 &&~~0.05/0.13 &&~~0.041/0.072 &\cr
&\hfil		     - 10 && 0.111 && 0.05 &&~~0.06/0.20 &&~~0.046/0.092 &\cr
&\hfil		     - 12 && 0.053 && 0.05 && 0.06 && 0.045 &\cr
&\hfil		     - 12 && 0.176 && 0.30 && 1.20 && 0.067 &\cr
&Truss et al. (2001) \hfil && 0.150 && 0.26 && 0.29 && 0.037 &\cr
&Foulkes et al. (2004) \hfil && 0.100 && $^*$)&& 0.10/0.15 && 0.081/0.103 &\cr
&\hfil			     && 0.100 && 0.05 && 0.13/0.18 && 0.072/0.087 &\cr
&\hfil			     && 0.100 && 0.10 && 0.23/0.31 && 0.069/0.082 &\cr
&\hfil			     && 0.100 && 0.20 && 0.56/0.85 && 0.071/0.090 &\cr
&&&&&&&&&&\cr
\noalign {\hrule height 0.3truemm}
}}$$
\hskip 25truemm $^*)~~~~R_i=R_{WD}~$ (full disk).
\baselineskip=14truept plus 1pt minus 1pt
\parskip=12truept
\end{table}

In spite of those problems we decided to use the 2D light curves taken 
from Hirose and Osaki (1990, Fig.8), Murray (1996, Fig.11), Murray (1998, 
Figs.6 and 7), Truss et al. (2001, Fig.7) and Foulkes et al. (2004, Fig.2). 
Results are listed in Table 1, where the fourth column 
gives the bolometric amplitudes $a_{bol}$ obtained from those light 
curves, while the last column -- the mean value of $A_V$ obtained from 
individual values corresponding to the four different values of $R_{sh}$ 
(see above). In all cases when two values of $a_{bol}$ and $A_V$ are listed, 
the first of them refers to the smooth part of the light curve, while 
the second -- to the narrow peak. 

The resulting visual amplitudes are generally smaller than $A_V\sim 0.10$. 
The only exception is the first entry representing one of the early 
models by Hirose and Osaki (1990). 
Excluding this case and using only values of $A_V$ corresponding to 
the smooth parts of model light curves we get $<A_V>=0.064$. 
Worth noting are the last four entries based on light curves from 
Foulkes et al (2004, Fig.2) which provide a useful test of our assumptions 
and calculations. The first of them refers to the full disk, while the 
three others -- to outer rings with different values of $R_i$.  
The bolometric amplitudes differ by factor of $\sim 5$. The resulting 
visual amplitudes, however, are practically identical.

% ***Table 2
\begin{table}[h!]
\parskip=0truept
\baselineskip=0pt
\bigskip
\centerline{Table 2}
\medskip
\centerline{ Amplitudes of Superhumps from 3D Simulations }
\medskip
$$\offinterlineskip \tabskip=0pt
\vbox {\halign {\strut
\vrule width 0.3truemm #&     %1
\quad#\quad&		      %2  mod
\vrule#&		      %3
\quad\hfil#\hfil\quad&	      %6  q
\vrule#&		      %7
\quad\hfil#\hfil\quad&	      %8  Ri/A
\vrule#&		      %9
\quad\hfil#\hfil\quad&	    %10 Abol
\vrule#&		      %11
\quad\hfil#\hfil\quad&	    %12 Av
\vrule width 0.3 truemm # \cr %13
\noalign {\hrule height 0.3truemm}
&&&&&&&&&&\cr
& Model/Run \hfil && $q$ && $R_i/A$ && $a_{bol}$ && $A_V$ &\cr
&&&&&&&&&&\cr
\noalign {\hrule}
&&&&&&&&&&\cr
&Smith et al. (2007) \hfil  - 5 && 0.220 && 0.30 && 0.25 && 0.028 &\cr
&\hfil			    - 6 && 0.212 && 0.30 && 0.29 && 0.030 &\cr
&\hfil			    - 7 && 0.176 && 0.30 && 0.40 && 0.037 &\cr
&\hfil			    - 8 && 0.143 && 0.30 && 0.45 && 0.039 &\cr
&\hfil			    - 9 && 0.111 && 0.30 && 0.39 && 0.036 &\cr
&\hfil			   - 10 && 0.081 && 0.30 && 0.48 && 0.043 &\cr
&&&&&&&&&&\cr
\noalign {\hrule height 0.3truemm}
}}$$
\baselineskip=14truept plus 1pt minus 1pt
\parskip=12truept
\end{table}

We now turn to the light curves published by Smith et al. (2007, Fig.4).  
They were based on their 3D simulations and calculated with respect to 
the luminosity of the outer ring with $R_i=0.3A$. 
Results are listed in Table 2. 
As can be seen, the visual amplitudes are very small, their 
average value $<A_V>=0.036$ being 10 times smaller than $<a_{bol}>$.  
This difference results primarily from the fact that $a_{bol}$ did not 
include contribution from the inner parts of the disk, responsible for 
a large fraction of the total bolometric luminosity and only for a small 
part of the visual luminosity.

% ***section 4
\section { Discussion }

Results presented above can be summarized as follows: 
Even in the case of 2D simulations, producing bolometric light curves 
with strongly overestimated amplitudes, we find that -- in spite of that 
-- the corresponding visual amplitudes are very small. 
Using more reliable bolometric light curves produced by 3D simulations 
we obtain visual amplitudes which are only $<A_V>=0.036$. 
This is $\sim 10$ times too low to explain the typical amplitudes 
of superhumps $A_V\sim 0.3$. 
In view of this discrepancy we must conclude that the tidal model 
for superhumps fails to explain their amplitudes.

\bigskip

\centerline {\bf References }

\bigskip

\refitem Foulkes, S.B., Haswell, C.A., Murray, J.R., Rolfe, D.J. 
         2004 \MNRAS {349} {1179}

\refitem Hirose, M., Osaki, Y. 1990 \PASJ {42} {135}

\refitem Hellier, C. 2001 {\it Cataclysmic Variable Stars} (Springer).  

\refitem Murray, J.R. 1996 \MNRAS {279} {402}

\refitem Murray, J.R. 1998 \MNRAS {297} {323}

\refitem Simpson, J.C., Wood, M.A., Burke, C.J. 1998 \BA {7} {255}

\refitem Smak, J. 2008 \Acta {58} {55}

\refitem Smith, A.J., Haswell, C.A., Murray, J.R., Truss, M.R., 
         Foulkes, S.B. 2007 \MNRAS {378} {785}

\refitem Truss, M.R., Murray, J.R., Wynn, G.A. 2001 \MNRAS {324} {L1}

\refitem Warner, B. 1995 {\it Cataclysmic Variable Stars} 
         (Cambridge University Press). 

\refitem Whitehurst, R. 1988 \MNRAS {232} {35} 

\end{document}